 \documentclass[112pt]{elsarticle}

 \usepackage{epsfig}
\usepackage{multirow}
\usepackage{amssymb}

\journal{Physics Letters B}

\begin{document}

\begin{frontmatter}




\title{Evidence of a new state in $^{11}$Be observed in the $^{11}$Li $\beta$-decay}

 \author[csic]{M. Madurga}
 \author[csic]{M.J.G. Borge\corref{cor}}
 \author[csic]{M. Alcorta}
 \author[complutense]{L.M. Fraile}
 \author[aarhus]{H.O.U. Fynbo}
 \author[goteborg]{B. Jonson}
 \author[aarhus]{O. Kirsebom}
 \author[gsi]{G. Mart\'inez-Pinedo}
 \author[goteborg]{T. Nilsson}
 \author[goteborg]{G. Nyman}
 \author[csic]{A. Perea}
  \author[uam]{A. Poves}
 \author[aarhus]{K. Riisager}
 \author[csic]{O. Tengblad}
 \author[goteborg]{E. Tengborn}
 \author[isolde]{J. Van der Walle} 
 
 \cortext[cor]{Corresponding author.}

 \address[csic]{Instituto de Estructura de la Materia, CSIC, Serrrano 113bis, E-28006 Madrid, Spain}
 \address[complutense]{Departamento de F\'isica At\'omica, Molecular y Nuclear,  Universidad  Complutense, E-28040 Madrid, Spain} 
 \address[aarhus]{Department of Physics and Astronomy, University of Aarhus, DK-8000 {\AA}rhus, Denmark}
 \address[goteborg]{Fundamental Physics, Chalmers University of Technology, S-41296 G\"oteborg, Sweden}
 \address[gsi]{Gesellschaft f\"ur Schwerionenforschung, D-64291 Darmstadt, Germany}
 \address[uam]{Departamento de F\'isica Te\'orica, Universidad Aut\'onoma de Madrid, E-28049 Madrid, Spain}
 \address[isolde]{PH Department, CERN, CH-1211 Gen\`eve, Switzerland}



\begin{abstract}
{Coincidences between charged particles emitted in the $\beta$-decay  of $^{11}$Li were observed  using highly segmented detectors. The breakup channels involving three particles were studied in full kinematics allowing for the reconstruction of the excitation energy of the  $^{11}$Be states participating in the decay. In particular, the contribution of a previously unobserved state  at 16.3 MeV  in $^{11}$Be has been identified selecting the $\alpha$+$^7$He$\rightarrow\alpha$+$^6$He+n channel. The angular correlations between the $\alpha$ particle and the center of mass of the $^6$He+n system favors spin and parity assignment of 3/2$^-$ for this state as well as for the previously known state at 18 MeV.  }
\end{abstract}

\end{frontmatter}


The study of the $\beta$-decay of $^{11}$Li constitutes a challenge for experimentalists. The high Q$_{\beta}$ value (20.557(4) MeV \cite{smith,bachelet}), combined with the low particle separation energies in $^{11}$Be,  opens a plethora of decay channels, $\beta$-$\gamma$ \cite{roeckl}, $\beta$-n \cite{roeckl}, $\beta$-2n \cite{azumab2n}, $\beta$-3n \cite{azuma}, $\beta$-$\alpha$ \cite{madurga08:1}, $\beta$-t \cite{langevin84} and $\beta$-d \cite{mukha}. These decay channels involve a variety of emitted particles over a wide energy range, which only recently has been open for detailed spectroscopic studies due to the availability of segmented detectors. The $\beta$-n channel, corresponding to the lowest energy threshold, at 0.504(6) MeV, has recently been studied in a series of $\beta$-$\gamma$-n coincidence experiments \cite{morrissey,hirayama} and  $\beta$-delayed $\gamma$ doppler broadening studies \cite{fynbo04,sarazin}, which mapped the  states of $^{11}$Be fed in the $^{11}$Li $\beta$-decay up to 10.6 MeV excitation energy. Information about the spin and parity of some of these states has also been obtained \cite{hirayama}. The remaining $\beta$-delayed channels include the fragmentation of $^{11}$Be states into more complex channels, in particular the five body 2$\alpha$3n channel makes the identification of the original state in $^{11}$Be difficult.

The first $\beta$-delayed charged particle emission study done in coincidences and with time of flight particle identification  established the presence of two different decay channels, three-body n$\alpha^6$He and five-body 2$\alpha$3n \cite{langevin81}. Two states in $^{11}$Be were proposed from energy considerations to contribute to these decays, one at 10.59(5) MeV identified in reaction studies \cite{selover} and a previously unobserved state proposed at $\sim$18.5 MeV \cite{langevin81}. The second state was later confirmed by an experiment using a gas-Si telescope for particle identification \cite{borge}. The energy spectra of the  $^{9,10}$Be, $^{4,6}$He and \emph{d},\emph{t} decay products were studied and the recommended energy for the state was  18.15(15) MeV. This state will be referred from here on as the 18 MeV state.

Furthermore, there have been experimental hints of $\beta^-$ feeding to states in between these two states in $^{11}$Be. In the gas-Si telescope experiment it was reported \cite{borge} that a peak at 1.2 MeV in the $^{9}$Be$+$$^{10}$Be  spectrum could be interpreted either as neutron emission from a $^{11}$Be 14.5 MeV state to the ground state in $^{10}$Be, or as neutron emission from the $^{11}$Be 18 MeV  state to the first excited state in $^{10}$Be. A later experiment \cite{fynbo04} that studied the neutron-recoil doppler broadening of $\gamma$-lines following $^{11}$Li decay reported a 6\% of $\beta$-feeding to excited states in $^{11}$Be in the region  between the 10.6 and 18 MeV states. 

Additional information on the $^{11}$Be states above the charged particle thresholds has been obtained by two recent reaction studies \cite{bohlen97,ohnishi}. From the spin-parity assignment given to these new  states they are not expected to be fed in the $\beta$-decay of $^{11}$Li.

On the theoretical side, shell model calculations \cite{suz97} predict that a significant amount of the $\beta$-decay strength lays in the region between 10 and 18 MeV in $^{11}$Be. Certainly, the shape of the Gamow-Teller strength distribution depends on the s/p mixing of the $^{11}$Li ground state wave function. 

These results underline the importance of studying the $^{11}$Li $\beta$-delayed charged particle emission in great detail.  In this Letter we present the study of  the $^{11}$Be excited states, fed in the $^{11}$Li $\beta$-decay,  using the  \linebreak $^{11}$Be$^*$$\rightarrow$$\alpha$+$^7$He$\rightarrow$n+$\alpha$+$^6$He channel, previously proposed in \cite{madurga08:1}.  We have established the presence of this channel in the breakup of the $^{11}$Be 18 MeV state \cite{langevin81,borge}. Complete kinematical analysis is possible from the detection of the energy and direction of the emitted particles provided by the use of segmented detectors, which allows to reconstruct the $^7$He+$\alpha$ channel to obtain the excitation energy spectrum of $^{11}$Be. Here we find the contribution of a previously unidentified state at 16 MeV, and determined, from the angular correlation study, the spin and parity of both states.


The experiment was carried out at the ISOLDE PSB Facility at CERN, where the $^{11}$Li activity was produced in fragmentation reactions in a Ta target irradiated with 1.4 GeV protons from the CERN PS Booster. The target container was connected to a surface ionization source and the produced ions were extracted with a voltage of 30 kV. The  1$^+$ $^{11}$Li ions were mass separated from the different isotopes  using the general purpose separator (GPS). The $^{11}$Li beam was subsequently brought to the center of the setup where it was stopped in a 50 $\mu$g/cm$^2$ carbon foil. The use of a thin foil, which minimizes the energy loss of the emitted charged particles, was possible due to the low extraction voltage used in the experiment. The production in our experimental chamber was estimated to be 800 $^{11}$Li ions/s, obtained from the daughter $^{11}$Be  $\beta \alpha$ activity \cite{alburger}. 

The experimental setup consisted of three particle telescopes, with 60 $\mu$m thick Double Sided Silicon Strip Detectors (DSSSD's) backed by Si pads for $\beta$ detection. Each DSSSD has its charge collection divided in 16 vertical and 16 horizontal strips, obtaining an angular resolution of $\sim$3.5$^o$ per 3$\times$3 mm$^2$ pixel.  The telescopes were mounted in fixed positions on the surface of an aluminum table. The two detectors facing each other covered solid angles of 4\% and 6\% of 4$\pi$, respectively \cite{madurga09}. 

The silicon detectors were calibrated using $^{148}$Gd and the standard triple alpha source ($^{239}$Pu, $^{241}$Am and $^{244}$Cm). An analysis method developed within our collaboration \cite{uffed} takes advantage of the highly segmented nature of the DSSSD's to precisely define the path of the detected particle and thus allowing for the reconstruction of the energy losses in the non-active layers. The combined effect of the dead layers and electronic noise limits the detection threshold to 160 keV for alpha particles.

\begin{figure*}
\begin{center}
\includegraphics[width=8cm]{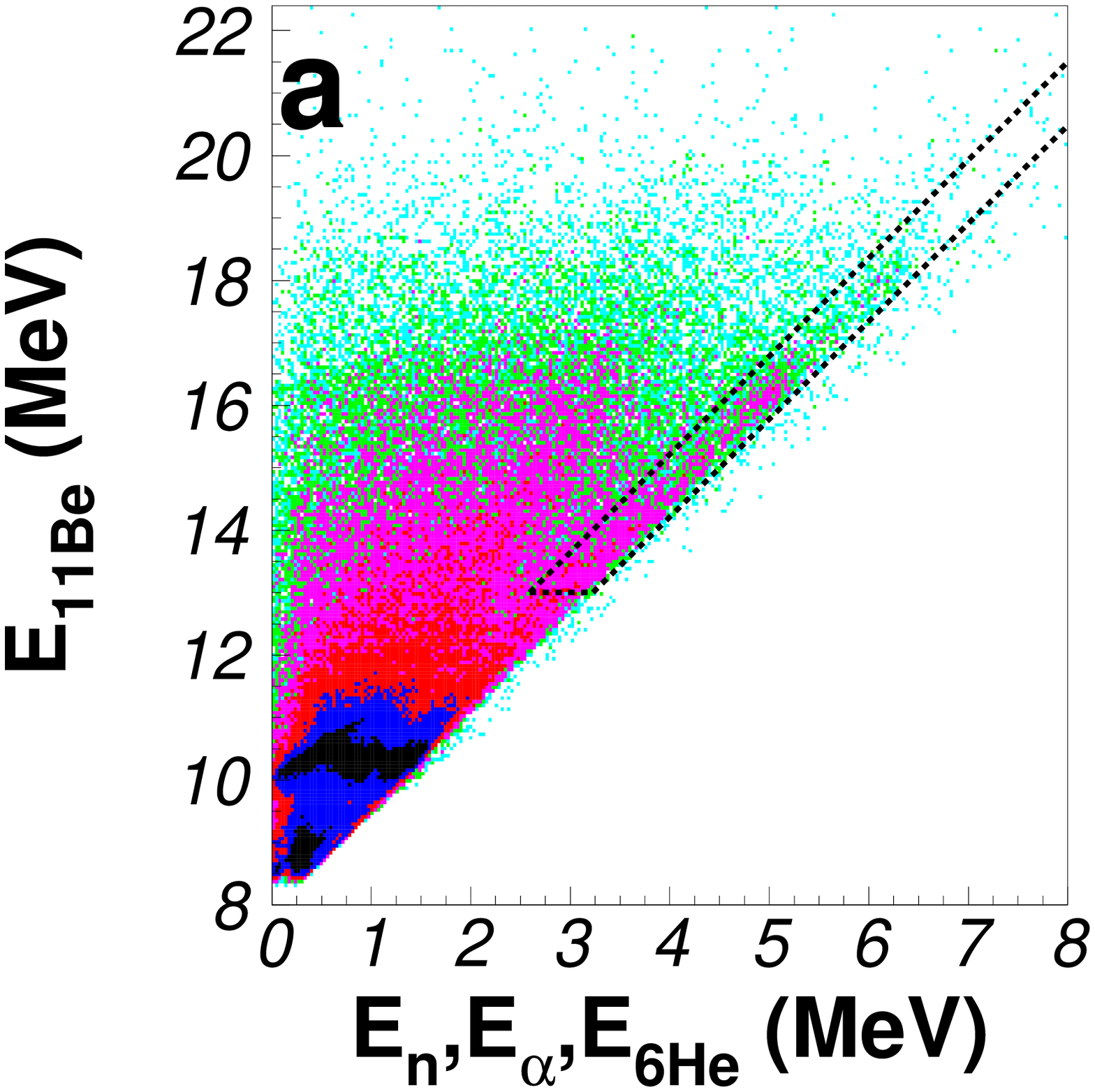}\includegraphics[width=8cm]{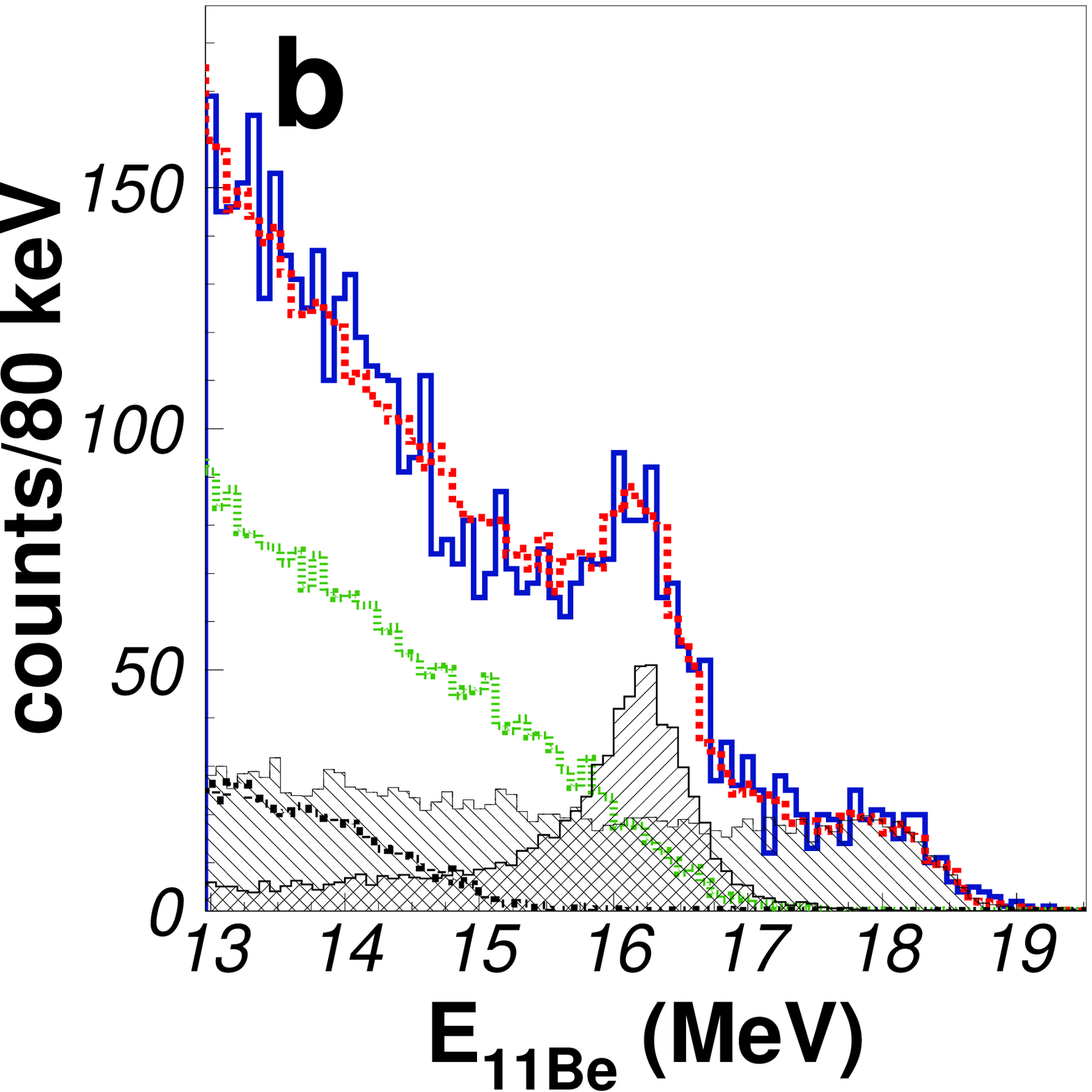}
\caption{(color online) Left: (a) scatter plot of the individual  $\alpha$, $^6$He and (reconstructed) neutron energies plotted versus the $^{11}$Be excitation energy, recorded within the first 60 ms after the proton impact. The charged particles were measured in opposite detectors, and the energy of the neutron reconstructed assuming 3-body breakup. The line of 10 MeV offset and 11/10 slope corresponds to the neutron emission of the $^{11}$Be 10.6 MeV state \cite{madurga08:1,langevin81}. The two dotted lines indicate the gate around the 11/7 interval corresponding to breakup through the $^7$He(gs) resonance. Right: (b) reconstructed $^{11}$Be excitation energy spectrum corresponding to the events inside the interval shown in the left panel. The good fit obtained in the Monte-Carlo simulation of the breakup process is shown in red dashed line over the data. The $^7$He+$\alpha$ break up contribution of the $^{11}$Be states at 16 and 18 MeV is shown with hatched area histograms, and the black dash-dotted (bellow the hatched area) and  green dotted lines correspond to the 2$\alpha$3n breakup of the for $^{11}$Be 16 and 18 MeV states respectively .\label{fig1} }
\end{center}
 
\end{figure*}

We concentrate here on the charged particle coincidences recorded in opposite detectors, setting the angle between the detected particles between 120$^o$ and 180$^o$. This geometry is optimum for the study of three body processes by detection of at least two of the products in coincidences. In our case 90\% of the coincidences were recorded in  the pair of opposite detectors. The time window for the analysis was set to the first 60 ms after the proton impact, time long enough to allow for the diffusion of the products from the target. This time window enhances the observation of the $^{11}$Li $\beta$-decay over the decay of the descendants. The directions of the detected charged particles were determined from pixel identification, which allowed for the calculation of their momenta as well as the deposited energy. In case of a three body process one can reconstruct the energy of the third undetected particle, the neutron. Assuming that the particle of lowest energy is the most massive one, the individual neutron, $\alpha$ and $^6$He energies are plotted versus excitation energy in $^{11}$Be in Fig. \ref{fig1}(a). The advantage of this scatter plot, for sequential breakup of broad states,  is that the first emitted particle will be along a line whose slope and offset are given by the mass ratio of the fragment to $^{11}$Be and the Q-value of the breakup, respectively \cite{fynbo00}. In the plot most of the events are $\alpha$-$\alpha$ coincidences from the  five body 2$\alpha$3n channel \cite{langevin81,madurga08:1}, which are not correctly assigned in this plot as the three neutrons are not detected. Therefore, the  events associated with the 2$\alpha$3n channel are distributed over the plot and act as background for the other channels. There are two distinct features in Fig. \ref{fig1}(a). First, a horizontal line at 10.6 MeV excitation energy, associated to the breakup of the 5/2$^-$ 10.6 MeV state in $^{11}$Be \cite{hirayama}. Second, a grouping along the line of 11/7 slope and with 8.35 MeV offset. In this type of scatter plot, in case of sequential breakup through an intermediate resonance of energy Q$_R$ and mass M$_R$, the energy of the first emitted particle E$_1$ will appear along a diagonal line following E=Q$_R$+$\frac{M}{M_R}$E$_1$, where M and E are the mass and excitation energy of the decaying state. In our case Q$_R$ corresponds to the $^7$He+$\alpha$ separation energy in $^{11}$Be, Q$_R$=8.35(2) MeV \cite{bachelet,audi}. Therefore,  the line in Fig. \ref{fig1}(a) corresponds to $\alpha$ particles emitted in the breakup of $^{11}$Be states through the ground state of the $^{7}$He resonance \cite{madurga08:1}. There are two regions with accumulations of events on this line, around 16 MeV and 18 MeV excitation energy, indicating the presence of states in $^{11}$Be at these energies.

We selected events in the $^{11}$Be$^*$$\rightarrow$$^7$He(gs)+$\alpha$$\rightarrow$$^6$He+$\alpha$+n channel by gating on the charged particle with the highest energy, which is assumed to be the $\alpha$ particle. The two dotted lines in Fig.~\ref{fig1}(a) show the interval analyzed, between the lines of slope of 11/7 and offsets of 8.93 and 7.93 MeV. The $^{11}$Be excitation energy spectrum for these events, obtained from the sum of n, $\alpha$ and $^{6}$He energy plus Q$_{n\alpha^6He}$,  is shown in Fig. \ref{fig1}(b). The two groupings of events observed in the scatter plot of Fig. \ref{fig1}(a) correspond to the clear peak around 16 MeV, plus the contribution at higher energies of the previously observed  $^{11}$Be  state at 18 MeV  \cite{langevin81,borge}. The peak at around 16 MeV indicates the presence of a state in $^{11}$Be  previously unreported in the literature. 

We performed a Monte-Carlo simulation of the breakup  of the 16 and 18 MeV states in $^{11}$Be into the three body n$\alpha^6$He and five body 2$\alpha$3n  channels, the latter channel considered as a broad background. The resonances were described using the single-channel non-interfering R-matrix formalism as described in the Appendix A of Ref. \cite{nyman}. The $\alpha$ particle penetrability was calculated using the orbital angular momentum values compatible with the spin-parity of the $^{11}$Be states and the 3/2$^-$ ground state in $^7$He. The penetrability of the neutron connecting the $^7$He ground state with the 0$^+$ ground state in $^6$He was calculated for l=1 orbital angular momentum. The five body 2$\alpha$3n breakup for the two states was modeled following the prescription given in Ref. \cite{madurga08:1}. The simulation including the contribution of the different channels is shown as the red histogram of Fig. \ref{fig1}(b), with each component color-coded following the legend. The parameters used to describe the different resonances are listed in Table \ref{table1}.

The resonance parameters were evaluated via $\chi^2$ tests of the $^{11}$Be excitation energy spectrum. We performed a series of Monte-Carlo simulations varying the energy centroid and the reduced width individually, fitting the resulting $\chi^2$ values to a parabola to obtain the minimum. We obtained E$_0$=16.3(1) MeV and $\gamma^2$=0.05(1) MeV, with a full width at half maximum (FWHM) of 0.7(1) MeV, for the first state and E$_0$=18.4(3) MeV and $\gamma^2$=0.11(4), FWHM=1.6(6) MeV, for the second. For both resonances the channel radius was set to 4.9 fm, r=r$_0$(A$_1^{1/3}$+A$_2^{1/3}$) with r$_0$=1.4 fm, and the resonance shift parameter was calculated using the formulation of the single-channel non-interfering R-matrix formalism as described in the Appendix A of Ref. \cite{nyman}.  The energy and width of the latter state were previously determined in \cite{borge} from the fit of singles spectra. Values of E$_{0}$=18.2(2) MeV and FWHM=1.25(15) MeV were obtained from the $^{9,10}$Be spectrum; E$_0$=18.0(1) MeV and FWHM=0.8(1) MeV from the $^{4,6}$He spectrum and E$_0$=18.15(15); and FWHM=1.3(3) from the triton spectrum. The resonance parameters obtained in this work coincide within error bars with the previously published parameters, although are more in line with the values obtained from the $^{9,10}$Be spectrum.  Analysis of the triton spectrum from the same experiment yields an energy and width of this state of 18.35(30) and 1.5(4) MeV respectively \cite{madurga09}, consistent with the results obtained from the n$\alpha^6$He channel presented here.

\begin{figure}
\begin{center}
 \includegraphics[width=7.5cm]{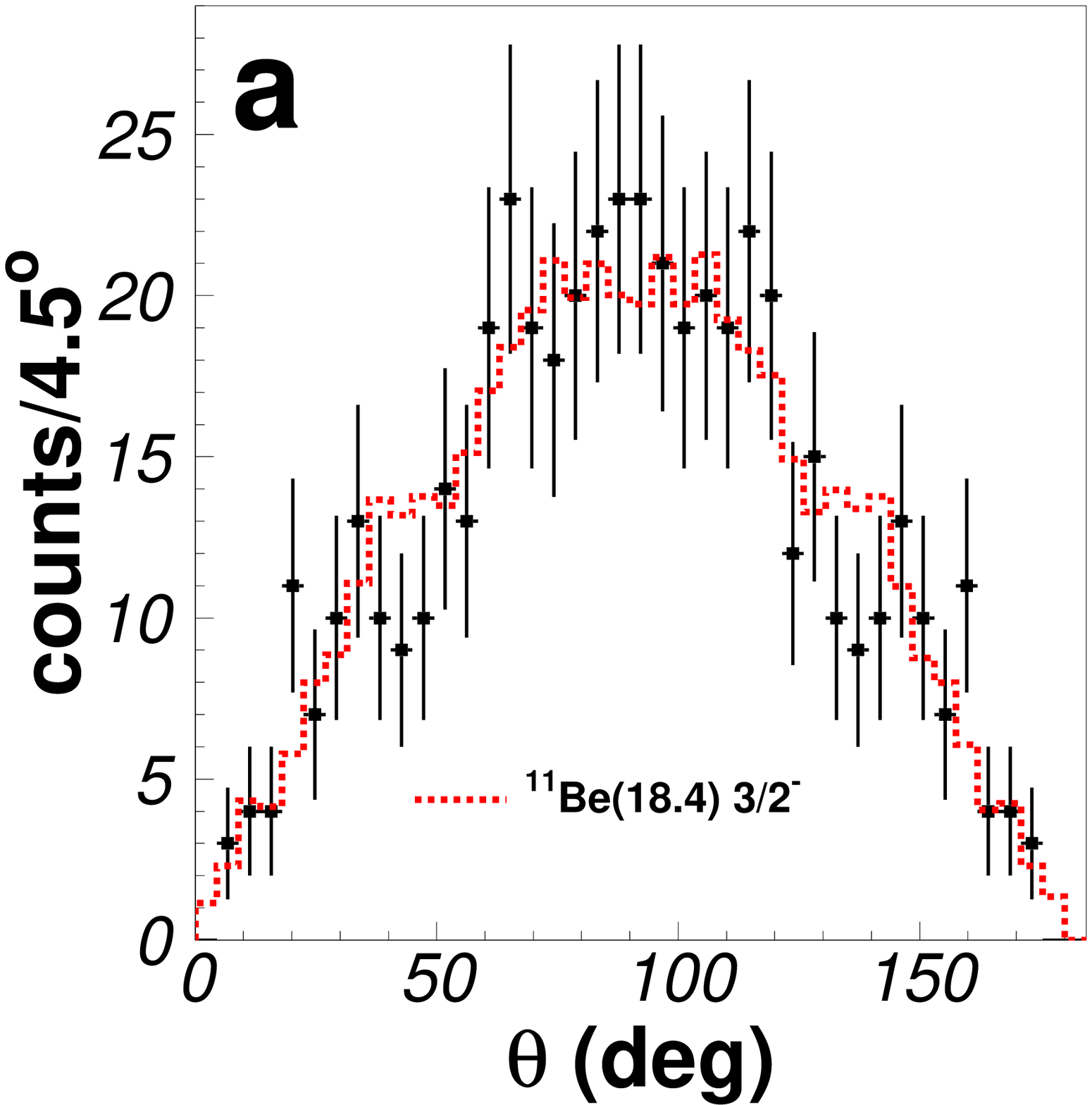}\includegraphics[width=7.5cm]{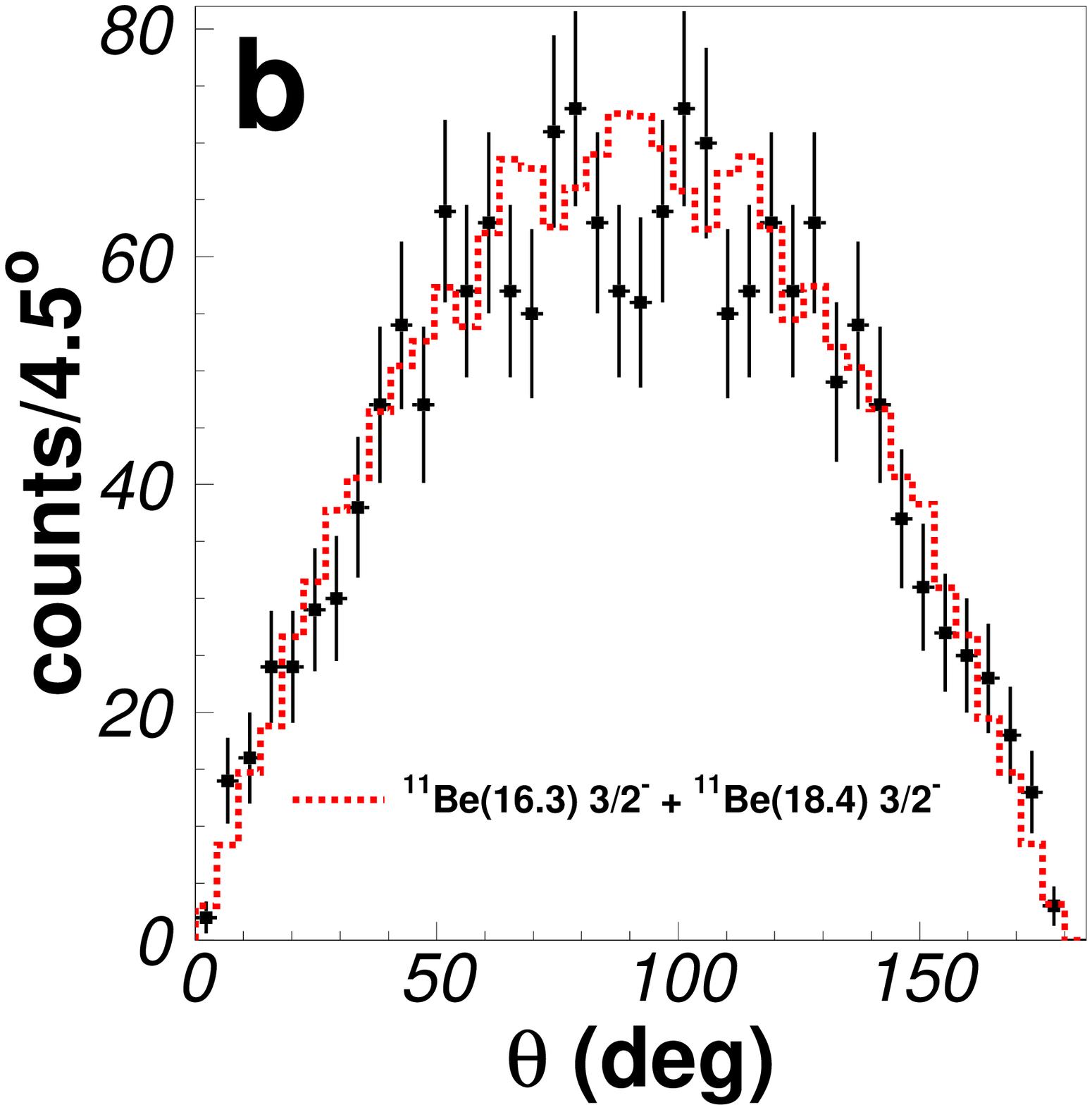}
\caption{(Color online) Left: (a) the angular distribution $\theta$ for events gated on the $^{11}$Be 18 MeV state is shown. The Monte-Carlo simulation of the most favored spin and parity assignments is overlaid. The red dashed histogram corresponds to spin J=3/2 for the $^{11}$Be 18 MeV state, which is the favored spin value from the $\chi^2$ test. Right: (b) angular distribution for events gated on the $^{11}$Be 16 MeV state. Again, the most favored Monte-Carlo simulation is shown, in this case J=3/2 (red dashed histogram) for both states at 16 and 18 MeV in $^{11}$Be.  \label{fig2}}
\end{center}
\end{figure}

The fact that the three body breakup channel is sequential and occurs through an intermediate state, $^7$He, with J$\neq$0,1/2 permit the use of the angular correlations to deduce the spin of the initial state in $^{11}$Be. The angle $\theta$ between these two directions is distributed according to the expression 1+A$_2$(2$\cos^2{\theta}$-1)/2, where A$_2$ depends on the spin of the initial and intermediate resonances. As the $^{11}$Li ground state is 3/2$^-$ \cite{selove}, the selection rules for  allowed transitions indicate that mainly 1/2$^-$, 3/2$^-$ and 5/2$^-$ states in $^{11}$Be are populated, whereas the spin and parity of the intermediate $^7$He(gs) resonance is 3/2$^-$ \cite{tilley}. We calculated the appropriate A$_2$ parameters from the parameterization given in \cite{biederharn}, obtaining A$_2$=1, 0 and -0.714 for 1/2, 3/2 and 5/2 respectively,   and performed a $\chi^2$ test of the simulation of the angular distribution of the breakup assuming every possible spin. In order to separate the contribution of each individual level we gated the sum energy spectrum in the following way: (i) the 15.7 to 16.6 MeV interval is chosen for the characterization of the 16 MeV state, as its contribution in this region is higher than the contribution of the 18 MeV state, see Fig. \ref{fig1}(b), (ii) the gate for the 18 MeV state was chosen  between 17.4 and 18.6 MeV, asymmetric with respect to the known resonance centroid  due to the weight of the Fermi factor. The experimental angular distributions for both cases are shown in Figs. \ref{fig2}(b) and (a) respectively, with the simulations for the most favored spin  assignment overlaid, and color coded according to the legend. One has to keep in mind that the spin of the 18 MeV state had to be studied first, as part of the statistics in the region of the 16 MeV state comes from the tail of the 18 MeV state. The results of the $\chi^2$ test, for 19 degrees of freedom,  for the 18 MeV state were $\chi^2$=~67.54 for 1/2$^-$, 16.62 for 3/2$^-$ and 30.62 for 5/2$^-$, which show that a spin of 3/2$^-$ is favored. In the case of the 16 MeV state, the results of the $\chi^2$ tests, setting the spin of the 18 MeV state to 3/2$^-$, again for 19 degrees of freedom, are $\chi^2$=~70.50 for 1/2$^-$, 20.57 for 3/2$^-$ and 58.68 for 5/2$^-$, leaving 3/2$^-$ as the only possible spin. In summary, the analysis of angular correlations in the $^7$He(gs)+$\alpha$ channel supports an assignment of spin of 3/2$^-$ for both states in $^{11}$Be.


The $\beta$-decay branching ratios to the $^{11}$Be states discussed in this Letter were obtained from the Monte-Carlo simulations, rather than directly from the coincidence data. This is due to the difficulty in determining the experimental coincidence efficiency for the different channels discussed, as it depends on the kinematics of the channel in question. The Monte-Carlo simulation, including realistic detector solid angle coverages and energy thresholds, was matched to the observed intensity of the coincidences in opposite detectors, and the intensity of each channel was determined by the total initial intensity of the simulation before applying the experimental restrictions or the experimental conditions.  The branching ratios were obtained using for normalization the $\beta \alpha$ activity from the decay of the daughter nucleus $^{11}$Be (T$_{1/2}$=13.81(8)s \cite{selove}), measured without any time restrictions. The branching ratio of the $^{11}$Be $\beta\alpha$ channel following $^{11}$Li beta decay is obtained from the product of two terms, the  $^{11}$Li $\beta$-feeding to the first excited state in $^{11}$Be assuming that the $\beta$-feeding to the $^{11}$Be ground state is negligible, and the $^{11}$Be $\beta$-delayed branching ratio of the $^{11}$Be $\beta\alpha$ channel. The first term, the feeding to the $^{11}$Be 320 keV state is taken as the weighted average of the values from \cite{morrissey,mjborge,bjornstad,aoi,detraz}, resulting in  7.4(3)\%. The second term,  the $^{11}$Be $\beta\alpha$ channel has only been determined once before \cite{alburger}, with a value of 2.9(4)\%. The resulting branching ratios are shown in the last column of Table \ref{table1}. The sum of the branching ratios to the 16 and 18 MeV states, 0.31(5)\%,  is comparable to the previously published values of $\beta$-feeding to the 18 MeV  state  in $^{11}$Be, 0.30(5)\% \cite{langevin81} and 0.39(7)\% \cite{madurga08:1}.

\begin{figure}
\centerline{\epsfig{file=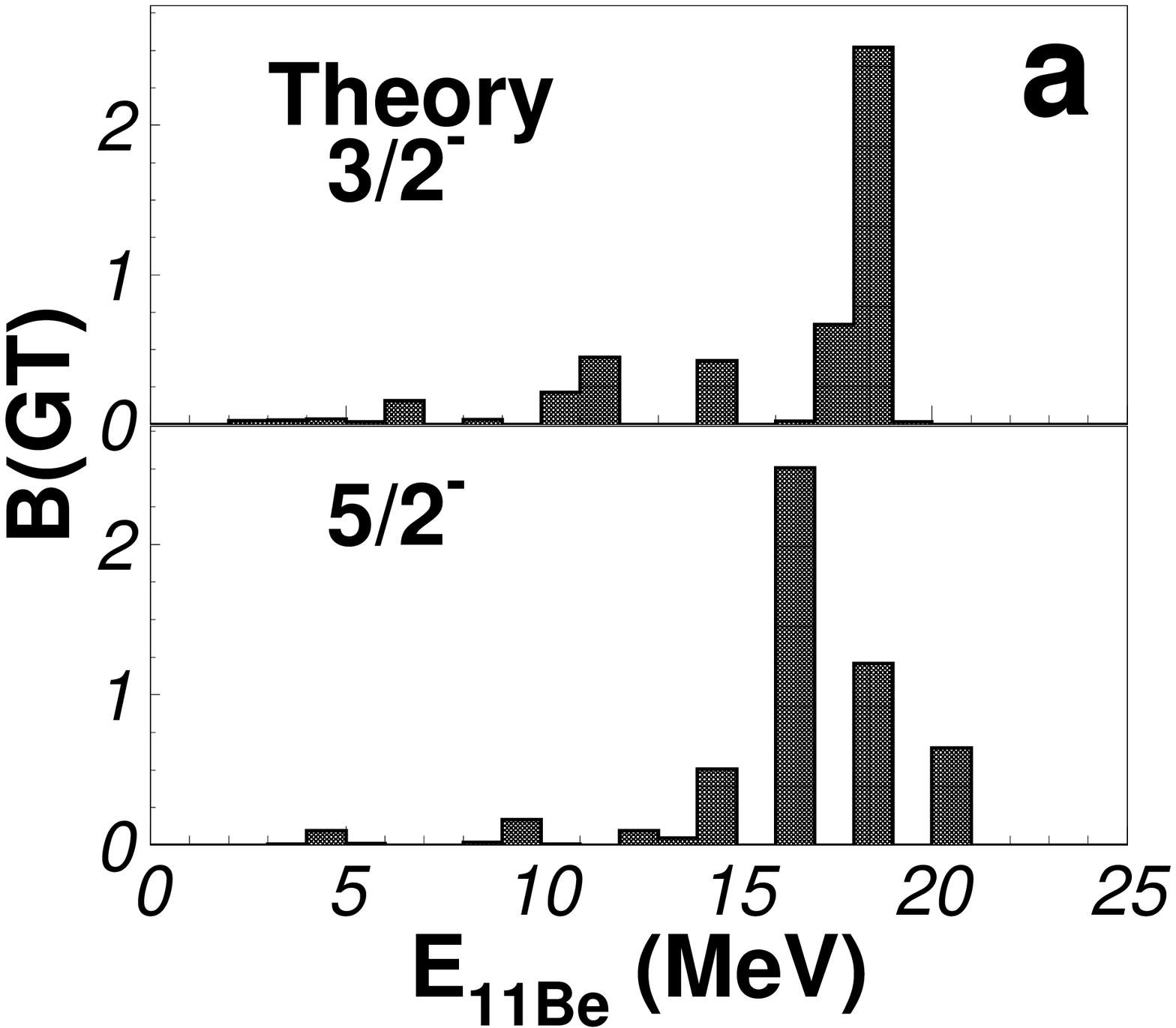,width=7cm}\epsfig{file=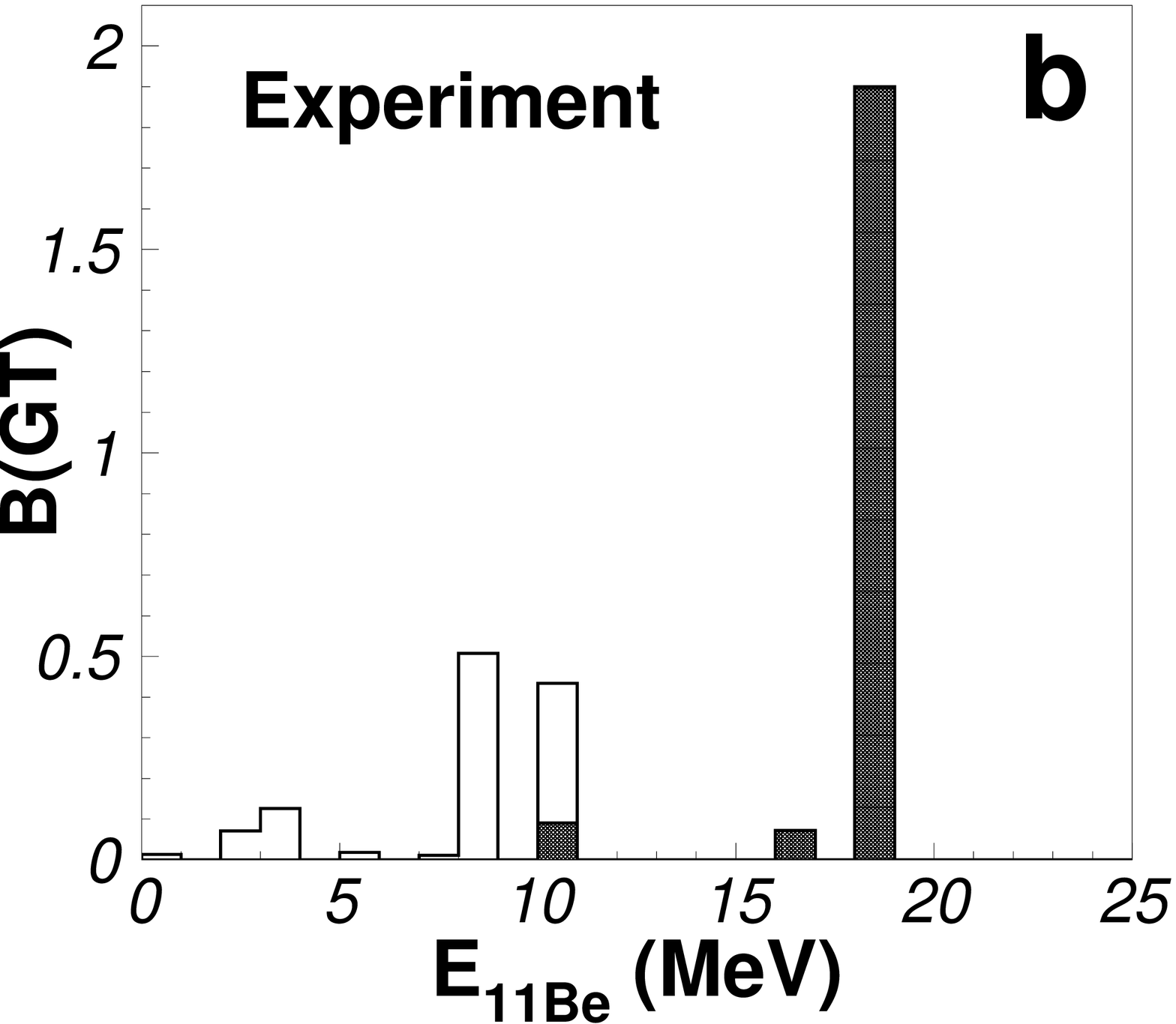,width=7cm}}

\caption{Left: (a) shows the B(GT) distribution to excited states in $^{11}$Be from  a shell model calculation using the unrestricted excitation space in the $p$ and $sd$ shells \cite{gabriel}. The upper panel  shows the B(GT) distribution for 3/2$^-$ states in $^{11}$Be.  The lower panel shows the distribution for 5/2$^-$ states in $^{11}$Be. Right: (b) shows the B(GT) transition values to excited states in $^{11}$Be is shown. The values obtained in this work are shown in solid bins. The values to states below the  charged particle emission thresholds are calculated from the branching ratios reported in \cite{fynbo04,hirayama} and assuming the levels are narrow and shown in open bins.  The overall shape of the calculation for 3/2$^-$ states agrees well with the experimental distribution, further supporting the 3/2$^-$ assignment discussed from the angular correlations. \label{fig3} }

\end{figure}

The B(GT) values are calculated from the branching ratios. This approach can be problematic for broad levels, as the B(GT) is normally defined as a factor in the R-matrix formalism (see \textit{e.g.} \cite{nyman,barker}). Instead, we used a modified \textit{ft}-value, where the Fermi function was substituted by its average over the state resonance shape, which was modeled using the single-level single-channel R-matrix formalism \cite{nyman}. In the case of the 18 MeV state in $^{11}$Be we included the neutron emission channels to states in $^{10}$Be, with branching ratios taken from \cite{borge}, and the $\beta$t value obtained in this experiment \cite{madurga09}. The resulting B(GT) values are shown in Fig. \ref{fig3}(b), represented as solid bins. The B(GT) values for levels below the charged particle emission thresholds are included for completion, and represented as open bins. These B(GT) values were calculated from the branching ratios of \cite{fynbo04,hirayama}, assuming the levels are narrow and using the regular \textit{ft}-value. The filled part of the 10 MeV bin in Fig. \ref{fig3}(b) shows the B(GT) deduced from the  branching ratio corresponding to charged particle emission from the 10.6 MeV state, taken from \cite{madurga08:1}.  

Fig. \ref{fig3}(a) shows the calculated B(GT) strength distribution to excited states in $^{11}$Be from shell model done using unrestricted excitation space in the $p$ and $sd$ shells (for details see \cite{mjborge,gabriel}). The inset panels in Fig. \ref{fig3}(a) shows the partial B(GT) distribution assuming feeding only to 3/2$^-$ or 5/2$^-$  states in $^{11}$Be, respectively. The shape of the calculated B(GT) distribution to 3/2$^-$ states at high excitation energy in $^{11}$Be (E$>$10.6 MeV) follows the trend of the experimental B(GT) values, supporting the 3/2$^-$ spin and parity assignment for these states given in this work. The experimental B(GT) strength for the 16 MeV state in $^{11}$Be is low compared to our shell model prediction and to the B(GT) strength calculated in \cite{suz97}. It should be stressed that the B(GT) values for the $^{11}$Be excited levels given in this work correspond to decays were at least two charged particles are involved in the breakup. However, one should be aware that there is previous evidence of a feeding of up to 6\% to levels in the region between 14-17 MeV excitation energy in $^{11}$Be decaying through neutron emission to low excited states in $^{10}$Be \cite{fynbo04}. This could explain the difference between the expected B(GT) distribution in this region and the experimental one. Unfortunately, with our setup, we just enhance the two-body and three-body decay channels involving at least two charged particles.

The study of the $^{11}$Be$^*$$\rightarrow$$\alpha$+$^7$He$\rightarrow$n+$\alpha$+$^6$He breakup channel presented in this Letter shows the power of the full kinematic analysis when three body channels participate in the decay. The presence of five body decay channels can obscure the signature of the three body ones, unless one has a suitable setup as the one used in this work. Full kinematics was achieved by detecting the $\alpha$ and the $^{6}$He in coincidence, and reconstructing the neutron energy and emission angle from energy and momentum conservation. This  allowed us to disentangle the presence of this three body channel in the decay of two excited states in $^{11}$Be, a known state at 18.4(3) MeV, and a state at 16.3(1) MeV, previously unreported in the literature. Although no experimental evidence of such state in $^{11}$Be existed, previous shell model calculations \cite{suz97,gabriel} predicted $\beta^-$ strength in this region. The study of the angular correlations in the breakup of these two states indicates a  3/2$^-$ spin assignment for both states.

This work has been supported by the Spanish Ministerio de Ciencia e Innovaci\'on (MICINN), under the projects FPA2007-62170 and CSD2007-00042, the European Union Sixth Framework through RII3-EURONS (contract no. 506065) and the Swedish Knut and Alice Wallenberg Foundation. M. Madurga acknowledges the support of the MICINN under the FPU program, FPU AP-2004-0002.  We acknowledge the help and support of the ISOLDE Collaboration during the experiment.



\clearpage

\begin{table}
\begin{center}
\caption{Level centroid and reduced widths  used in the R-Matrix description of the states modeled in the Monte-Carlo code. The width $\Gamma$ was obtained from a gaussian  fit of the R-matrix peak directly. The last column lists the branching ratios determined in this work following $^{11}$Li $\beta$-decay. The $^{11}$Li activity was deduced from the branching of the $\beta\alpha$ decay channel of the daughter $^{11}$Be. \label{table1}   }
\medskip
\begin{tabular}{ccccc||c}
  \hline \vspace{-7.5mm}  \\ 
 & E$_0$  & $\gamma^2$  & $\Gamma$ (FWHM) & Ref. & BR  \\
& (MeV) & (MeV) & (MeV) & & (\%) \\
\hline

\multirow{2}{*}{$^{11}$Be(16.3 MeV)} & \multirow{2}{*}{16.3(1)} & \multirow{2}{*}{0.05(1)} & \multirow{2}{*}{0.7(1)} & (3-body)~ this work & 0.006(1) \\
 & & & & (5-body)~ this work & 0.042(7) \\ \hline
\multirow{2}{*}{$^{11}$Be(18.4 MeV)} & 18.4(3) & 0.11(4) & 1.6(6) & (3-body)~ this work & 0.020(3)  \\ 
  & 18.15& 0.06 & 0.8 & (5-body)~ \cite{borge,madurga08:1} &  0.24(4) \\
\hline \vspace{-4.0mm}  \\
$^{7}$He(gs)  &  0.43$^a$ & 0.4 & 0.148(1) & \cite{tilley} \\
\hline
\vspace{-3.mm} \\
\multicolumn{6}{l}{$^a$ Above the $^6$He+$\alpha$ threshold.} \\
 \end{tabular}
\end{center}

\end{table}

\clearpage

\clearpage

\clearpage

\clearpage

\begin{thebibliography}{00}
\bibitem{smith}
M. Smith et al.,
Phys. Rev. Lett. 101  (2008) 202501. 

\bibitem{bachelet}
C. Bachelet et al.,
Eur. Phys. J. A25 s01 (2005) 31.
\bibitem{roeckl}
     E. Roeckl et al.,
    Phys. Rev. C10 (1974) 1181.

\bibitem{azumab2n}
    R.E. Azuma et al.,
    Phys. Lett. B43 (1979) 1652.

\bibitem{azuma}
    R.E. Azuma et al.,
    Phys. Lett. B96 (1980) 31.

\bibitem{madurga08:1}
M. Madurga et al.,
Nucl. Phys A810 (2008) 1.

\bibitem{langevin84}
    M. Langevin et al.,
    Phys Lett. B146 (1984) 176.

\bibitem{mukha}
    I. Mukha et al.,
    Phys. Lett. B367 (1996) 65.

\bibitem{morrissey}
D.J. Morrissey et al.,
Nucl. Phys. A627 (1997) 222.

\bibitem{hirayama}
    Y. Hirayama et al.,
    Phys. Lett. B611 (2005) 239.


\bibitem{fynbo04}
    H.O.U. Fynbo et al.,
    Nucl. Phys. A736 (2004) 39.

\bibitem{sarazin}
F. Sarazin et al.,
Phys. Rev. C70 (2004) 031302(R). 

\bibitem{langevin81}
    M. Langevin et al.,
    Nucl. Phys. A366 (1981) 449.

\bibitem{selover}
    F. Ajzenberg-Selove, E.R. Flynn and Ole Hansen,
    Phys. Rev. C17 (1978) 1283.


\bibitem{borge}
    M.J.G. Borge et al.,
    Nucl. Phys. A613 (1997) 199.

\bibitem{bohlen97}
H.G. Bohlen et al., Prog. Part. Nucl. Phys. 42 (1999) 17.

\bibitem{ohnishi}
T. Ohnishi et al., Nucl. Phys. A687 (2001) 38c.


\bibitem{suz97}
    T. Suzuki and T. Otsuka,
    Phys. Rev. C56 (1997) 847.

\bibitem{alburger}
    D.E. Alburger, D.J. Millener and D.H. Wilkinson,
    Phys. Rev. C23 (1981) 473.


\bibitem{madurga09}

M. Madurga et al.,
Eur. Phys. J. (2009), (online first), DOI 10.1140/epja/i2008-10769-0.

\bibitem{uffed}
    U.C. Bergmann, H.O.U. Fynbo and O. Tengblad,
    Nucl. Instr. and Meth. A515 (2003) 657.

\bibitem{fynbo00}
H.O.U. Fynbo et al.,
Nucl. Phys. A677 (2000) 38.

\bibitem{audi}
G. Audi, A.H. Wapstra and C. Thibault,
Nucl. Phys. A729 (2003) 337.


\bibitem{nyman}
    G. Nyman et al.,
    Nucl. Phys. A510 (1990) 189.

 


\bibitem{selove}
    F. Ajzenberg-Selove,
    Nucl. Phys. A506 (1990) 1.

\bibitem{tilley}
    D.R. Tilley et al.,
    Nucl. Phys. A745 (2004) 155.

\bibitem{biederharn}
L.C. Biederharn and M.E. Rose, Rev. Mod. Phys. 25 (1953) 729.


\bibitem{mjborge}
    M. J. G. Borge et al.,
    Phys. Rev. C55 (1997) R8.

\bibitem{bjornstad}
    T. Bj\"{o}rnstad et al.,
    Nucl. Phys. A359 (1981) 1.

\bibitem{aoi}
    N. Aoi et al.,
    Nucl. Phys. A616 (1997) 181.


\bibitem{detraz}
    C. D\'{e}traz,
    J. de Phys. Lett. 41 (1980) 459.

\bibitem{barker}
F.C Barker and E.K. Warburton,
Nucl. Phys. A487 (1988) 269.

\bibitem{gabriel}
G. Mart\'{i}nez-Pinedo, PhD Thesis, Universidad Autonoma de Madrid, unpublished, 1995.







\end{thebibliography}
\end{document}